\documentclass{elsart1p}

\usepackage{amssymb}
\usepackage{graphics}
\def\lsim{\raise0.3ex\hbox{$<$\kern-0.75em\raise-1.1ex\hbox{$\sim$}}}
\def\gsim{\raise0.3ex\hbox{$>$\kern-0.75em\raise-1.1ex\hbox{$\sim$}}}
\def\muBQS{\mu_{B,Q,S}}

\begin{document}

\begin{frontmatter}



\title{QCD thermodynamics at zero and non-zero density}


\author{Christian Schmidt (for RBC-Bielefeld and HotQCD Collaborations)}

\address{Universit\"at Bielefeld, Fakult\"at f\"ur Physik, D-33615 Bielefeld, Germany.}

\begin{abstract}
We present results on the QCD equation of state, obtained with two
different improved dynamical staggered fermion actions and almost
physical quark masses. Lattice cut-off effects are discussed in detail
as results for three different lattice spacings are available now,
i.e. results have been obtained on lattices with temporal extent of
$N_\tau=4,6$ and 8. Furthermore we discuss the Taylor expansion
approach to non-zero baryon chemical potential by means of an expansion 
of the pressure. We use the expansion coefficients to calculate various
fluctuations and correlations among hadronic charges. 
We find that the correlations reproduce the qualitative behavior of the 
resonance gas model below $T_c$ and start to agree with the free gas predictions
for $T\gsim 1.5T_c$.
\end{abstract}

\end{frontmatter}

\section{Introduction}
A detailed and comprehensive understanding of the thermodynamics
of quarks and gluons, e.g. of the equation of state is most desirable
and of particular importance for the phenomenology of relativistic
heavy ion collisions.
Lattice regularized QCD simulations at non-zero temperatures have
been shown to be a very successful tool in analyzing the
non-perturbative features of the quark-gluon plasma. Driven by both,
the exponential growth of the computational power of recent
super-computers as well as by drastic algorithmic improvements one is
now able to simulate dynamical quarks and gluons on fine lattices with
almost physical masses.

\section{Lattice parameter and scale setting}
\label{lcp}
We perform calculations on lattices of extent $16^3\times 4$,
$24^3\times 6$, using the p4fat3 action \cite{EoS} and $32^3\times 8$
with both p4fat3 and asqtad actions. The latter calculations are still 
preliminary and are currently performed
by the HotQCD collaboration \cite{hotQCDeos}.  For the generation of
gauge configurations we use the exact RHMC algorithm \cite{RHMC}.  For
each finite temperature calculation we perform a corresponding zero
temperature calculation on a lattice of at least the size
$N_\sigma^4$, where $N_\sigma$ is the spatial extent of the finite
temperature lattices.

The simulations are done on a line of constant physics (LCP), i.e.
the quark masses are kept constant in physical units. In practice, this has been
obtained by tuning the bare quark masses such that the meson masses of e.g.
pion, kaon and pseudo-scalar strange meson $\bar ss$ stay
constant in the QCD vacuum as we change the value of the coupling.
The strange quark mass was always fixed to its
physical value, by fixing kaon and $\bar s s$ to their
corresponding physical values \cite{EoS}. We find that the LCP can, to a good
approximation, be parameterized by a constant ratio of the bare
quark masses. Most calculations are done on a LCP which corresponds to
a pion mass of about 220 MeV. We do, however, also show preliminary
results with a light quark mass corresponding to about 150 MeV for the 
lightest pseudo-scalar mass \cite{soeldner}.


To set set the temperature scale in physical units, we determine two
distance scales, $r_0$ and $r_1$, from the zero temperature static
quark potential
\begin{equation}
\left(r^2\frac{{\rm d}V_{\bar q q}(r)}{{\rm d}r}\right)_{r=r_0}=1.65,
\qquad
\left(r^2\frac{{\rm d}V_{\bar q q}(r)}{{\rm d}r}\right)_{r=r_1}=1.0
\end{equation}
The ratio of both scales is only slightly quark mass dependent. It has
been determined in both discretization schemes consistently,
$r_0/r_1=1.4636(60)$ (p4fat3 \cite{EoS}) and 1.474(7)(18) (asqtad
\cite{asqtad}). The distance scales $r_0$ and $r_1$ have been related
to properties of the chamonium spectrum which allows to determine them
in physical units. We use here $r_0=0.469(7)$ fm as determined
in Ref. \cite{r0}. More details on the scale setting procedure, as well
as the parameterization of the LCP are given in Ref. \cite{EoS}.

\section{The equation of state}
\label{eos}
\begin{figure}
\begin{center}
\resizebox{0.575\textwidth}{!}{%
  \includegraphics{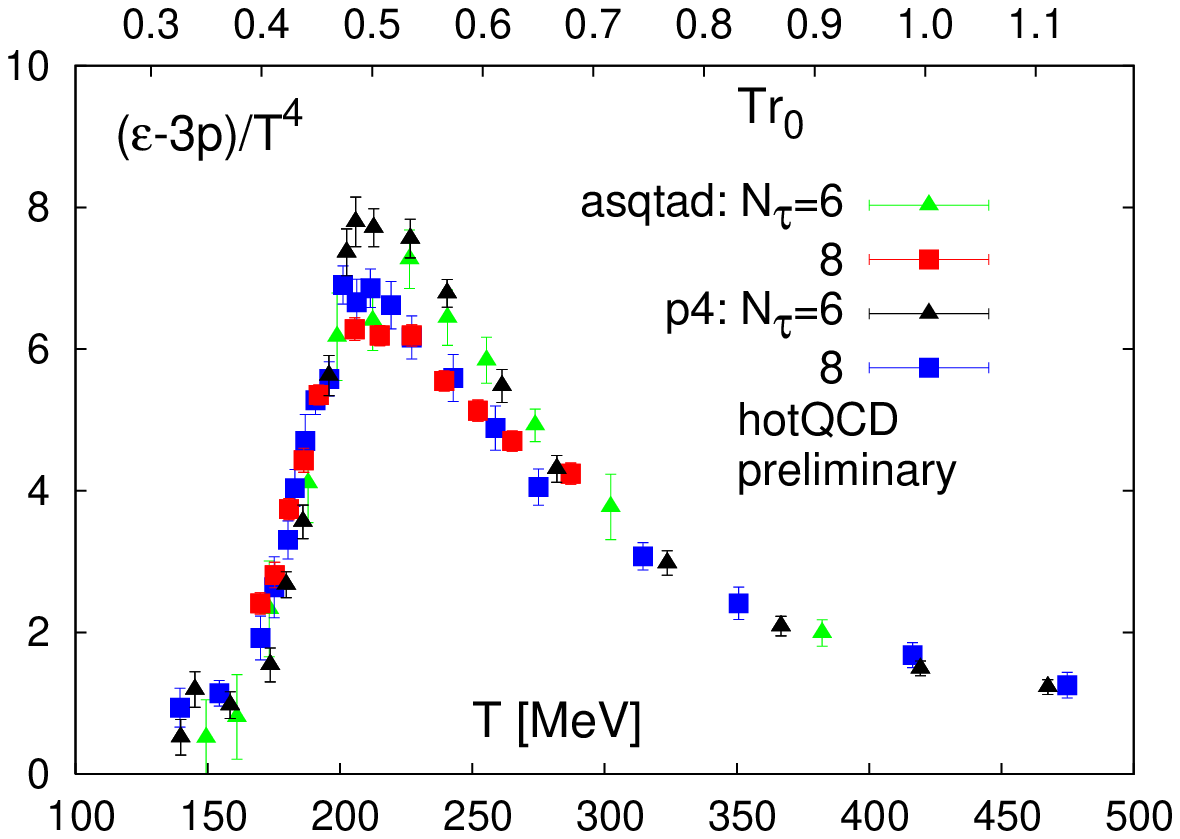}
}
\resizebox{0.4\textwidth}{!}{%
  \includegraphics{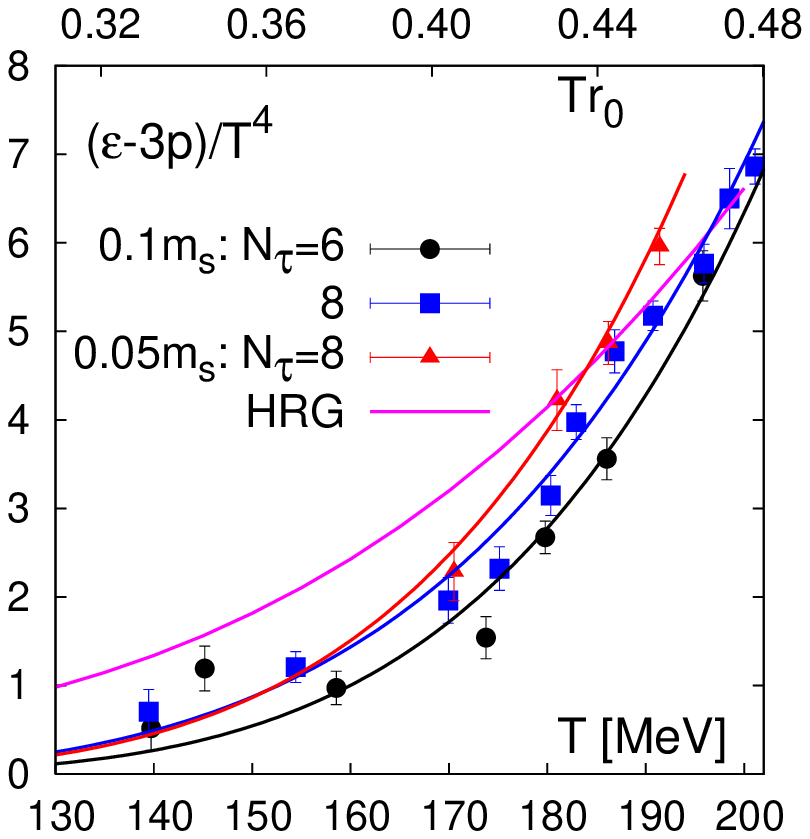}
}
\end{center}
\caption{ 
The trace anomaly, $(\epsilon -3p)/T^4$ on $N_\tau=6$ and $8$
lattices. On the left panel results are obtained with the p4fat3 and
asqtad actions, where $N_\tau=6$ results are from \cite{EoS} and
\cite{milc_eos} respectively, $N_\tau=8$ results are preliminary
hotQCD results \cite{hotQCDeos}. The right panel shows the low
temperature part of the trace anomaly in detail, here we plot only
results obtained by the p4fat3 action. Also shown on the right panel
are quadratic fits to the data, as well as the trace anomaly obtained
in the framework of the Resonance gas model. Light quark masses have
been constrained to be one tenth of the strange quark mass ($0.1m_s$), 
on the right panel we show, however, also preliminary results from
simulations with physical quark masses ($0.05m_s$) \cite{soeldner}.}
\label{fig:e-3p}  
\end{figure}

Along the line of constant physics, at sufficiently large volume and at zero chemical
potential, the temperature is the only intensive parameter that
controls the thermodynamics.  Consequently there exists only one
independent bulk thermodynamic observable that needs to be
calculated. All other quantities are than obtained by using standard
thermodynamic relations. On the lattice, it is convenient to first
calculate the trace anomaly in units of the fourth power of the
temperature, $\Theta^{\mu\mu}/T^4$. It is easily obtained as a
derivative of the pressure $p/T^4$, with respect to the temperature,
\begin{equation}
\frac{\Theta^{\mu\mu}}{T^4} \equiv \frac{\epsilon-3p}{T^4}
=T\frac{\partial}{\partial T}(p/T^4).
\label{eq:trace}
\end{equation} As the pressure is directly given by the partition
function, $p/T=V^{-1}\ln Z$, the calculation of the trace anomaly
requires only the evaluation of rather simple expectation
values. According to Eq. \ref{eq:trace} one then obtains the pressure
by
\begin{equation} \frac{p(T)}{T^4} - \frac{p(T_0)}{T_0^4} =
\int_{T_0}^{T} {\rm d}T' \frac{1}{T'^5} \Theta^{\mu\mu} (T') \;\; .
\label{pres}
\end{equation} Here $T_0$ is an arbitrary temperature value which
usually is chosen in the low temperature regime where the pressure and
other thermodynamic quantities are suppressed exponentially by
Boltzmann factors corresponding to the lightest hadronic states;
e.g. the pions.  The energy density is then obtained by combining
results for $p/T^4$ and $(\epsilon -3p)/T^4$, respectively.

In Fig.~\ref{fig:e-3p} (left) we show results for $\Theta^{\mu\mu}/T^4$
obtained with the asqtad and p4fat3 actions, respectively. The new
$N_\tau=8$ results \cite{hotQCDeos} are compared to $N_\tau=6$ results
taken from \cite{EoS,milc_eos}.  We note that the asqtad and p4fat3
formulations give results which are in good agreement with each
other. In fact, in quite a large temperature regime the agreement for
given lattice extent $N_\tau$ seems to be better than one could expect
in view of the overall cut-off dependence that is visible when
comparing results for $N_\tau=6$ and $N_\tau=8$ more closely.
They lead to a reduction of the peak height in $\Theta^{\mu\mu}/T^4$,
which is located at $T\simeq 200$~MeV and to a shift of the 
rapidly rising part of $\Theta^{\mu\mu}/T^4$ in the transition region to
smaller values of the temperature.  

In Fig.~\ref{fig:e-3p} (right) we show quadratic fits to
the data obtained by the p4fat3 action, to highlight the cut-off
effects. It is evident, that the $N_\tau=8$ data is shifted relative
to the $N_\tau=6$ data by about $8$~MeV at low temperatures, $T\simeq
160$~MeV. This shift decreases to about $5$~MeV at temperatures
$T\simeq 190$~MeV.

Light quark masses have been constrained to be one tenth of the
strange quark mass ($0.1m_s$), on the right panel we show, however,
also preliminary results from simulations with light quark masses
of $0.05m_s$. This again leads to a further shift in $\Theta^{\mu\mu}$
of approximately $5$~MeV at $T\simeq 190$~MeV towards lower
temperatures.

We also compare the results for $(\epsilon -3p)/T^4$ to results
obtained from the hadron resonance gas model. Details on the resonance
gas curve in Fig.~\ref{fig:e-3p} (right) will be given in
\cite{hotQCDeos}. The slope of $(\epsilon -3p)/T^4$ obtained by the
resonance gas model seems to be much smaller than the slope obtained
by the quadratic fits to the data.  Whether this points at deviations
of the equation of state at lower temperatures from resonance gas
behavior or is due to larger cut-off effects in the low temperature
regime requires further studies. We note that the lattice spacing
becomes larger at lower temperatures and violations of flavor
symmetry, which are inherent to the staggered fermion formulations at
finite lattice spacing, thus will become more important.

\begin{figure}
\begin{center}
\resizebox{0.48\textwidth}{!}{%
  \includegraphics{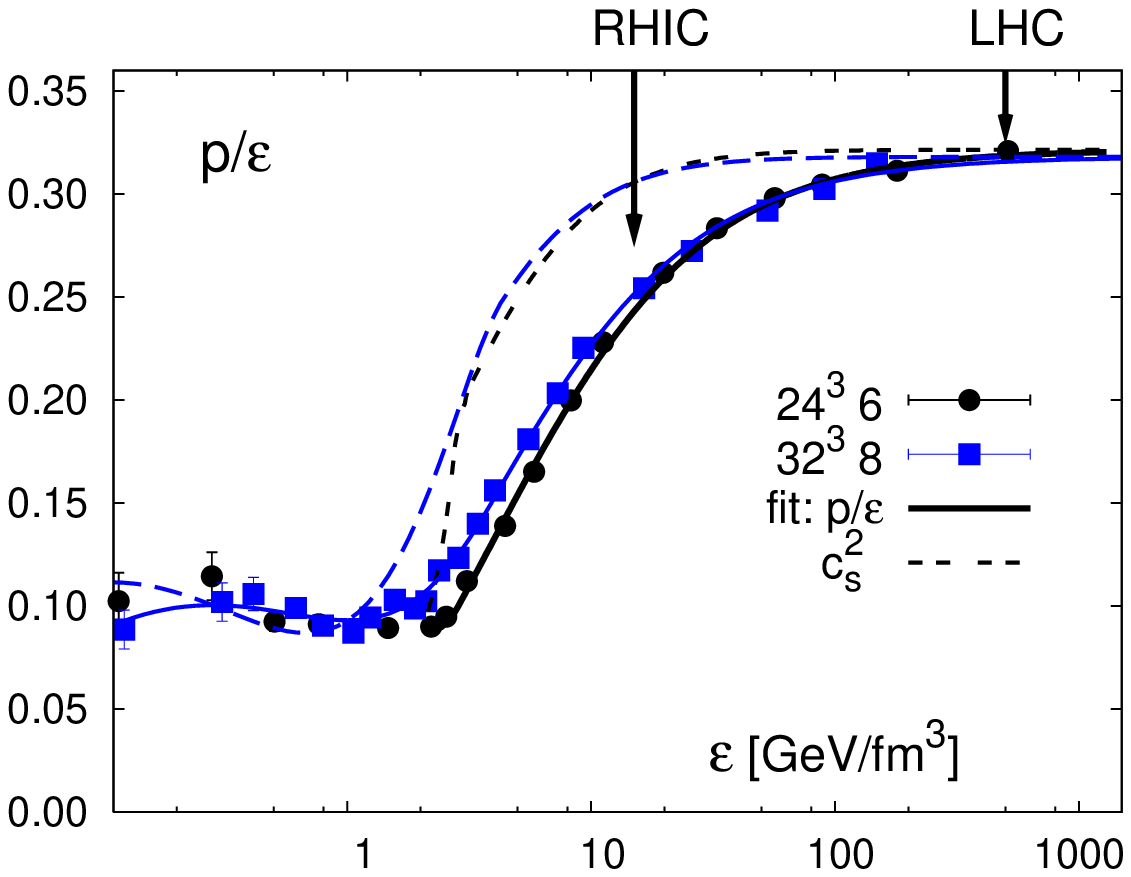}
}
\resizebox{0.48\textwidth}{!}{%
  \includegraphics{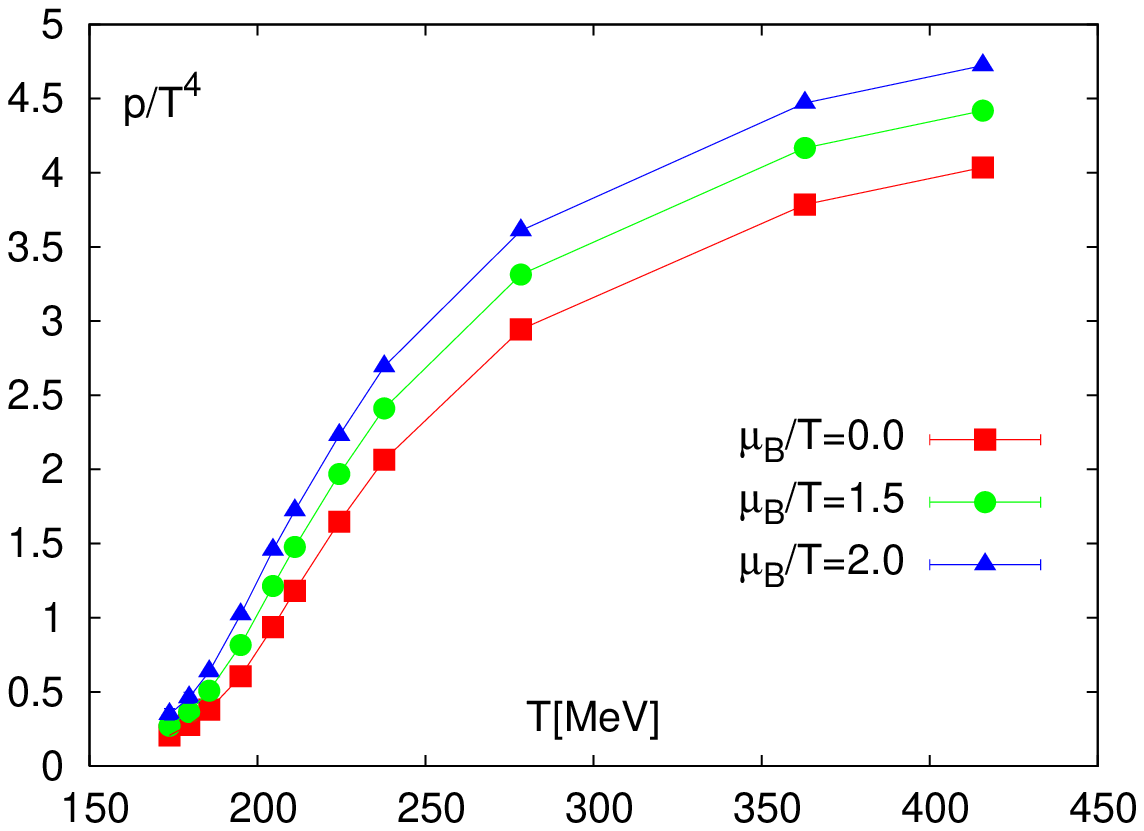}
}
\end{center}
\caption{The ratio of pressure and energy density as well as the
velocity of sound obtained in calculations with the p4fat3 action on
$N_\tau=6$ \cite{EoS} and $N_\tau=8$ \cite{hotQCDeos} lattices (left). 
The pressure in units of $T^4$ as funtion of temperature, for fixed and
non-zero values of $\mu_B/T$, obtained on $N_\tau=6$ lattices (right).}
\label{fig:eos}  
\end{figure}

The cut-off dependence observed in $\Theta^{\mu\mu}/T^4$ carries over
to the calculation of pressure and energy density; the former is
obtained by integrating over $\Theta^{\mu\mu}/T^5$ and the energy
density is then obtained by combining results for $p/T^4$ and
$(\epsilon -3p)/T^4$.  This is apparent in Fig.~\ref{fig:eos} (left) where we
show the ratio $p/\epsilon$ obtained with the p4fat3 action on
$N_\tau=6$ \cite{EoS} and $N_\tau=8$ \cite{hotQCDeos} lattices.
Cut-off effects are still visible in the vicinity of the 'softest
point' of the EoS, which is related to the peak position of
$(\epsilon-3p)/T^4$. We find that in the entire range of energy densities relevant
for the expansion of dense matter created at RHIC, $\epsilon \;
\lsim\; 10$~GeV/fm$^3$, the ratio $p/\epsilon$ deviates significantly
from the conformal, ideal gas value $p/\epsilon =1/3$.

This also is reflected in the behavior of the velocity of sound,
$c_s^2 = {\rm d}p/{\rm d}\epsilon$, which is shown in
Fig.~\ref{fig:eos} (left) by dashed lines. It starts deviating significantly
from the ideal gas value below $\epsilon \; \simeq\; 10$~GeV/fm$^3$
and reaches a value of about $0.1$ in the transition region at energy
densities $\epsilon\;\simeq\; 1$~GeV/fm$^3$.  Below the transition it
slightly rises again, but note, that for very small temperatures
$c_s^2$, as well as $p/\epsilon$ are sensitive the integration constant
$p_0(T_0)$ (see Eq.~\ref{eq:trace}). At present we have set $p_0(T_0)=0$,
for $T_0=100$~MeV.

\section{Non-zero chemical potential}
\label{density}
At non-zero chemical potential, lattice QCD is harmed by the
``sign-problem'', which makes direct lattice calculations with
standard Monte Carlo techniques at non-zero density practically
impossible.  However, for small values of the chemical potential, some
methods have been successfully used to extract information on the
dependence of thermodynamic quantitites on the chemical potential.
For an overview see, e.g. \cite{overview}.

We closely follow here the approach and notation used in
Ref.~\cite{eos6}. We start with a Taylor expansion for the pressure in
terms of the quark chemical potentials $\mu_{u,d,s}$, we obtain
\begin{equation}
\frac{p}{T^{4}}
=\sum_{i,j,k}c^{u,d,s}_{i,j,k}(T)\left(\frac{\mu_{u}}{T}\right)^{i}
\left(\frac{\mu_{d}}{T}\right)^{j}\left(\frac{\mu_{s}}{T}\right)^{k}.
\label{eq:PTaylor}
\end{equation}
The expansion coefficients $c^{u,d,s}_{i,j,k}(T)$ 
are computed on the lattice at zero chemical potential, using  
stochastic estimators. Some details are given in~\cite{details,latt08}.
We currently calculate the coefficients up to the 8th and 4th order,
on $N_\tau=4$ and $6$ lattices, respectively.
We find that
cut-off effects are small and of similar magnitude as those found for 
the trace anomaly $\Theta^{\mu\mu}$ \cite{latt08}. This was already
anticipated by the analysis of the cut-off corrections in the free gas
limit. Similar results for the asqtad action have been obtained 
in~\cite{milc_dens}.  

Alternatively to the quark chemical potentials one can introduce
chemical potentials for the conserved quantities baryon number $B$,
electric charge $Q$ and strangeness $S$ ($\mu_{B,Q,S}$), which are
related to $\mu_{u,d,s}$ via
\begin{equation}
\mu_{u}=\frac{1}{3}\mu_{B} +\frac{2}{3} \mu_{Q},\qquad
\mu_{d}=\frac{1}{3}\mu_{B}-\frac{1}{3}\mu_{Q},\qquad
\mu_{s}=\frac{1}{3}\mu_{B}-\frac{1}{3}\mu_{Q}-\mu_{S}.
\label{eq:chempot}
\end{equation}
By means of these relations the coefficients $c^{B,Q,S}_{i,j,k}$ of
the pressure expansion in terms of $\mu_{B,Q,S}$ are easily obtained,
in analogy to Eq.~\ref{eq:PTaylor}
\begin{equation}
\frac{p}{T^{4}}
=\sum_{i,j,k}c^{B,Q,S}_{i,j,k}(T)\left(\frac{\mu_{B}}{T}\right)^{i}
\left(\frac{\mu_{Q}}{T}\right)^{j}\left(\frac{\mu_{S}}{T}\right)^{k}.
\label{eq:PTaylor_hadronic}
\end{equation}
In Fig.~\ref{fig:eos} (right) we show the pressure expansion up to the 
fourth order in $\mu_B/T$. The results are obtained on $N_\tau=6$ lattices.
We find that corrections to the pressure arising from a non-zero chemical potential
are dominated by the second order expansion coefficient for moderate chemical
potentials of $\mu_B/T\lsim 2$ and are of the order of 10-20\% for $T>T_c$.

\section{Hadronic Fluctuations}
Quark number fluctuations are obtained from derivatives of the QCD
partition function with respect to the quark chemical potentials by the
fluctuation-dissipation theorem. The Taylor expansion coefficients
$c_{i,j,k}^{u,d,s}$, as defined in Eq.~\ref{eq:PTaylor}, can thus be
directly interpreted as quark number fluctuations at $\mu=0$.
However, quark fluctuations can not be detected directly in
experiments due to confinement. Therefore we will consider
fluctuations in terms of hadronic quantum numbers, i.e. baryon number
$B$, electric charge $Q$ and strangeness $S$, which are more easily
obtained by experiment. These fluctuations are related to the 
Taylor expansion coefficients $c_{i,j,k}^{B,Q,S}$, as given in
Eq.~\ref{eq:PTaylor_hadronic}. A recent overview on the physics of fluctuations 
in the context of heavy ion collisions was given in Ref.~\cite{koch}.
In general, the quadratic fluctuations $\chi_2^X$ at zero chemical
potentials can be obtained from the second order coefficient $c_2^X$
\begin{equation}
  c_2^X \equiv \left.\frac{1}{2VT^3}
    \frac{\partial^2 \ln Z}{\partial(\mu_X/T)^2}\right|_{\muBQS=0}
    = \frac{1}{2VT^3} \langle(\delta N_X)^2\rangle_0\ ,
\end{equation}
where $\delta N\equiv N-\langle N\rangle$ denotes the normalized net-density.
and
$\langle\ldots\rangle_0$ indicates that the expectation value has been
taken at $\muBQS=0$. Under such conditions, baryon number, electric
charge and strangeness vanish, and we have $\delta N=N$. We define 
quadratic and quartic charge fluctuations by
\begin{equation}
  \chi_2^X = \frac{1}{VT^3} \langle N_X^2\rangle_0 = 2c_2^X, \qquad
  \chi_4^X = \frac{1}{VT^3} \left( \langle N_X^4 \rangle_0
    - 3 \langle N_X^2 \rangle_0^2 \right) = 24c_4^X\ ,
\end{equation}
respectively, and correlations among two conserved charges by
\begin{eqnarray}
   \chi_{11}^{XY} &=& \frac{1}{VT^3} \left( \langle N_XN_Y \rangle_0 
     - \langle N_X \rangle_0 \langle N_Y \rangle_0 \right)
   	= c_{11}^{XY}\ ,
\end{eqnarray}
where $X,Y \in \{B,Q,S\}$.

\begin{figure}
\begin{center}
\resizebox{0.49\textwidth}{!}{%
  \includegraphics{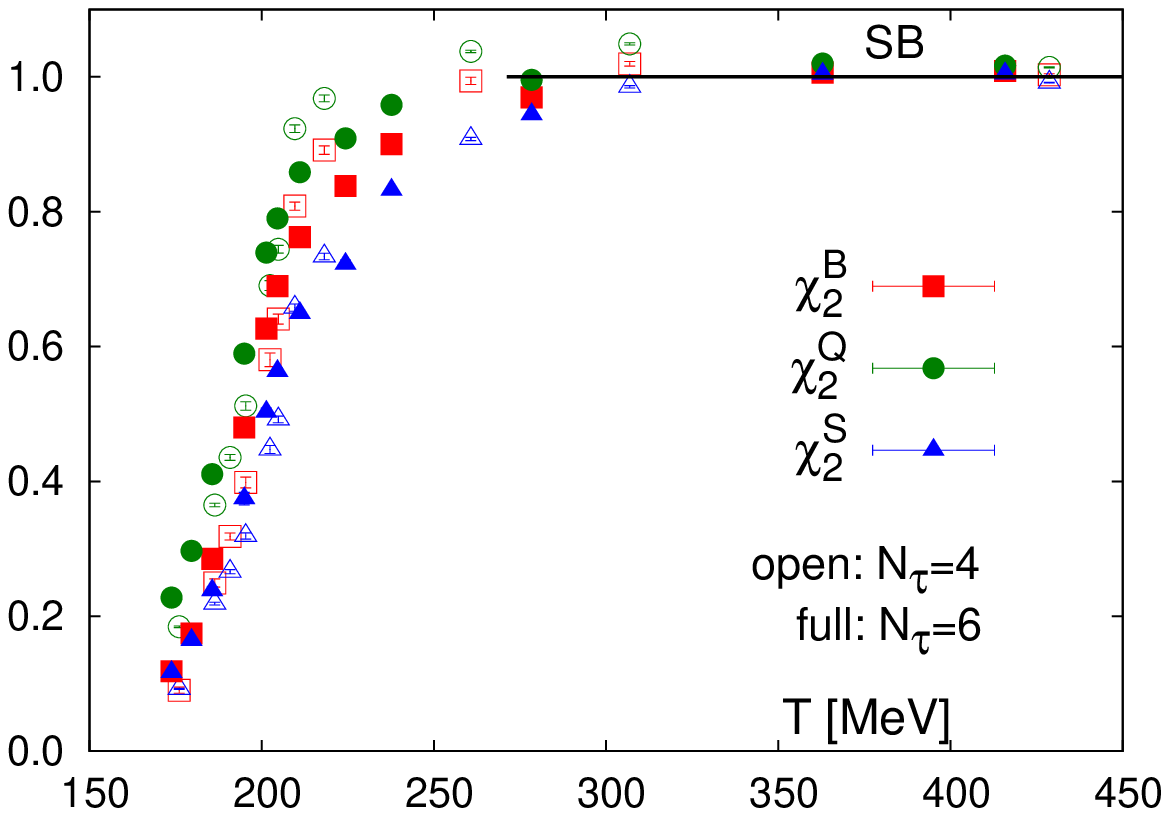}
}
\resizebox{0.49\textwidth}{!}{%
  \includegraphics{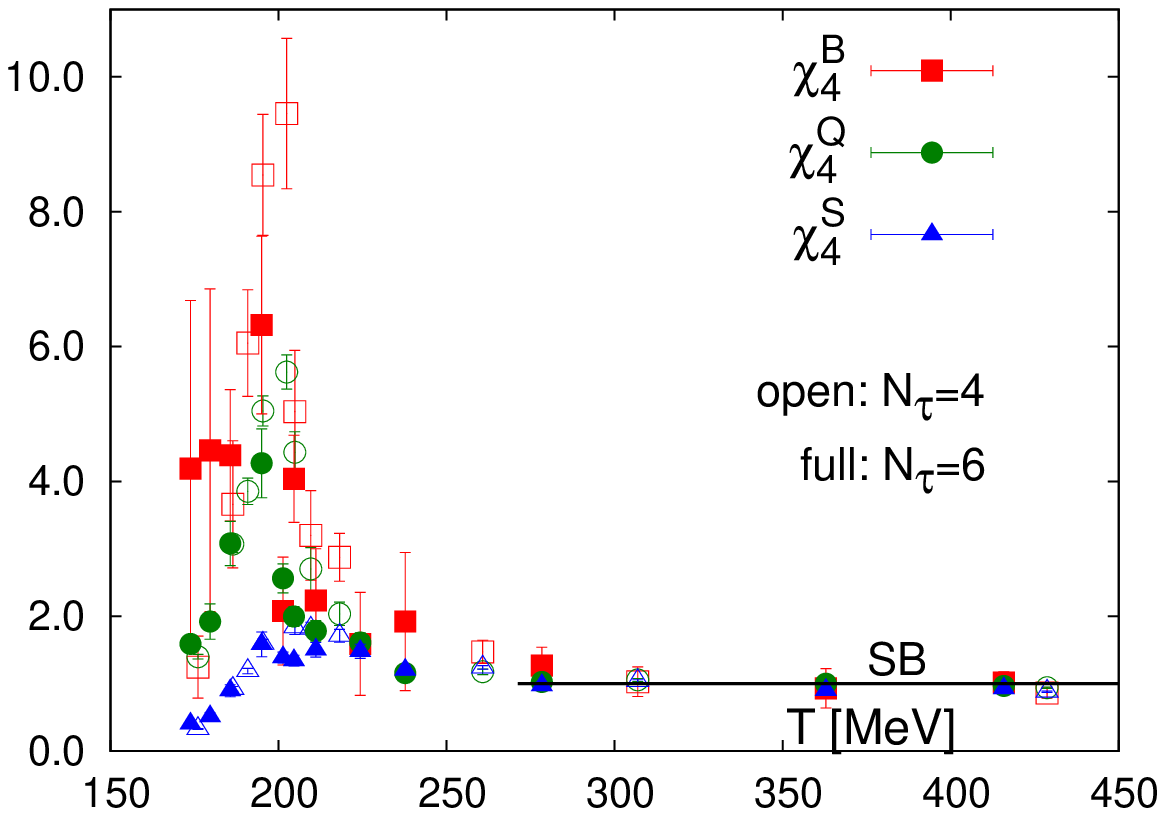}
}
\end{center}
\caption{
Quadratic and quartic fluctuations of baryon number, electric charge
and strangeness, normalized by their corresponding Stefan-Boltzmann value.
The results on $N_\tau=4$ lattices (open symbols)
and $N_\tau=6$ lattices (full symbols) are in good agreements.
}
\label{fig:c24BQS}  
\end{figure}
In Fig.~\ref{fig:c24BQS} we show results for quadratic and quartic 
fluctuations of $B$, $Q$ and $S$.  The quadratic fluctuations $\chi_2^{B,Q,S}$
rise rapidly in the transition region where the quartic fluctuations 
$\chi_4^{B,Q,S}$ show a peak. The peak height is more pronounced for the baryon
number fluctuations than for fluctuations of the strange quarks. 

We compare the results obtained on lattices with temporal extent $N_\tau = 4$
and $6$. We notice that they are in general compatible with each other, 
especially in the high temperature phase, where both quadratic and quartic
fluctuations approach the Stephan-Boltzmann limit quickly. The transition
temperature has been previously determined to be $T_c = 202$ MeV and $196$ MeV
on $N_\tau = 4$ and $6$ lattices respectively \cite{Tc}. We thus conclude that at
temperatures of about
$1.5T_c$ and higher, quadratic and quartic fluctuations of $B$, $Q$ and $S$
are well described by the ideal massless quark gas.

At low temperature, hadrons are the relevant degrees of freedom. 
The hadron resonance gas (HRG) model has been shown to provide a 
good description of thermal conditions at freeze-out. We thus compare
the fluctuations in the low temperature phase with a HRG model, where
we include all mesons and baryons with masses smaller than 2.5 GeV from 
the particle data book. 

\begin{figure}
\begin{center}
	\resizebox{0.32\textwidth}{!}{\includegraphics{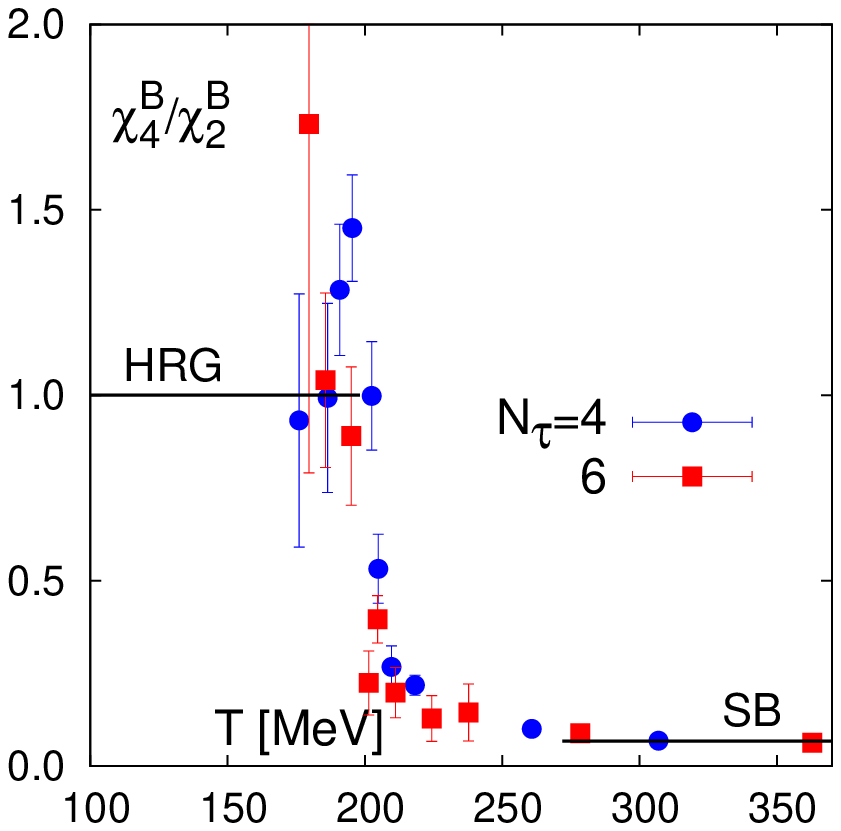}}
	\resizebox{0.32\textwidth}{!}{\includegraphics{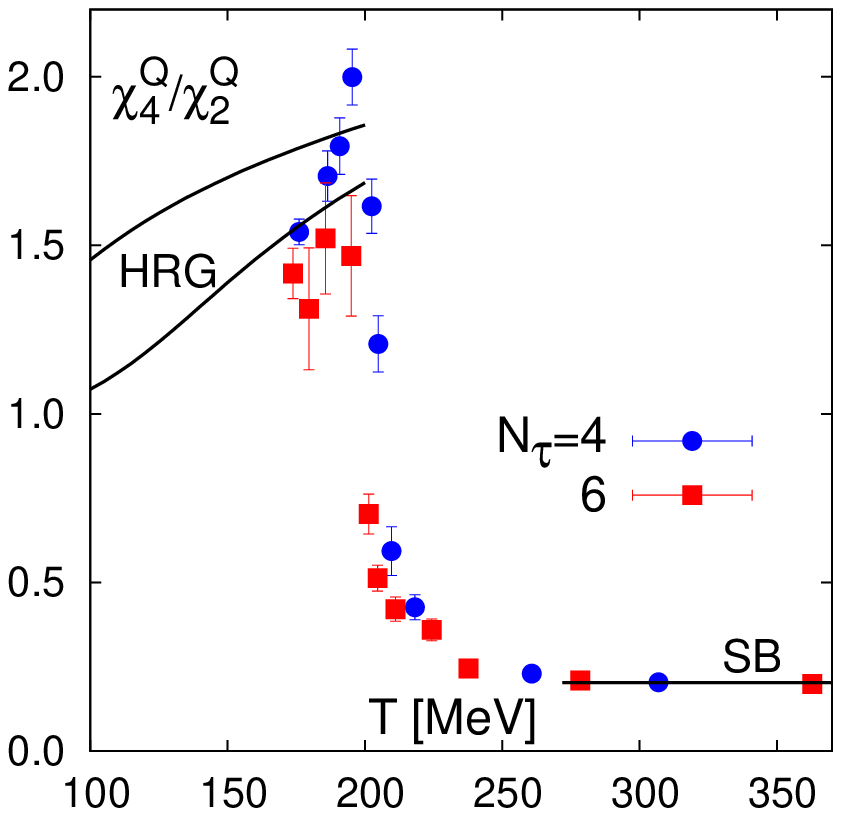}}
	\resizebox{0.32\textwidth}{!}{\includegraphics{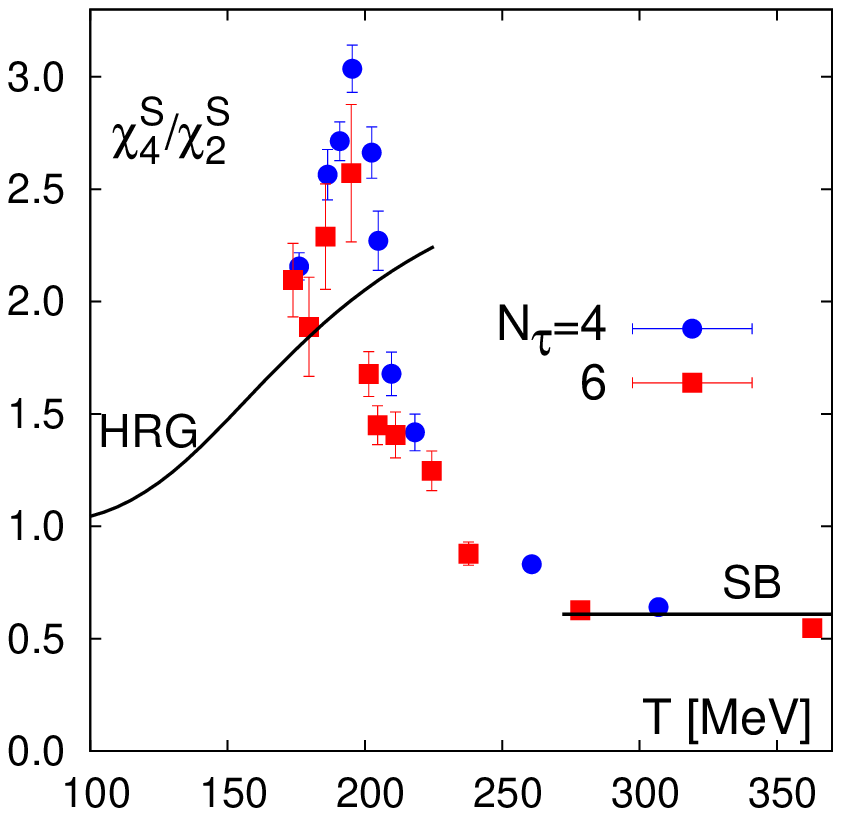}}
\end{center}
\caption{
Ratio of quartic and quadratic fluctuations of baryon number ($B$), strangeness ($S$)
and electric charge ($Q$). The two curves of the HRG model in the top right
figure correspond to charge fluctuations with physical pions (upper) and
infinitely heavy pions (lower curve).
}
\label{fig:R2BQS}
\end{figure}

In Fig.~\ref{fig:R2BQS}, we show the ratio of quartic and quadratic
fluctuations for $B$, $S$ and $Q$.  
In the HRG model, $\chi_4^B/\chi_2^B$ is 
easily obtained in the Boltzmann approximation, which is valid for 
a dilute baryonic gas in the temperature range of interest. One finds
that all details on the hadron mass spectrum and temperature dependence 
cancel and the result is a constant, given by the unit square of the 
baryonic charge (one for all baryons). This is
in fact reproduced by the lattice results shown in Fig.~\ref{fig:R2BQS} 
(top left).

The ratio of quartic and quadratic fluctuations
for $S$ and $Q$ are more complicated even in the
Boltzmann limit, since hadrons with different electric/strange charge
give rise to different contributions to the corresponding fluctuations. For
strangeness fluctuations, shown in Fig.~\ref{fig:R2BQS} (bottom), the Boltzmann
limit is still a good approximation; but for electric charge fluctuations,
the pion mass plays an important role. In order to check for the sensitivity
of electric charge fluctuations on the pion mass, we show in Fig.~\ref{fig:R2BQS}
(top right)
results of a HRG model calculation with physical pion masses and without
the pion sector, i.e. for infinitely heavy pions.

\begin{figure}
\begin{center}
	\resizebox{0.32\textwidth}{!}{\includegraphics{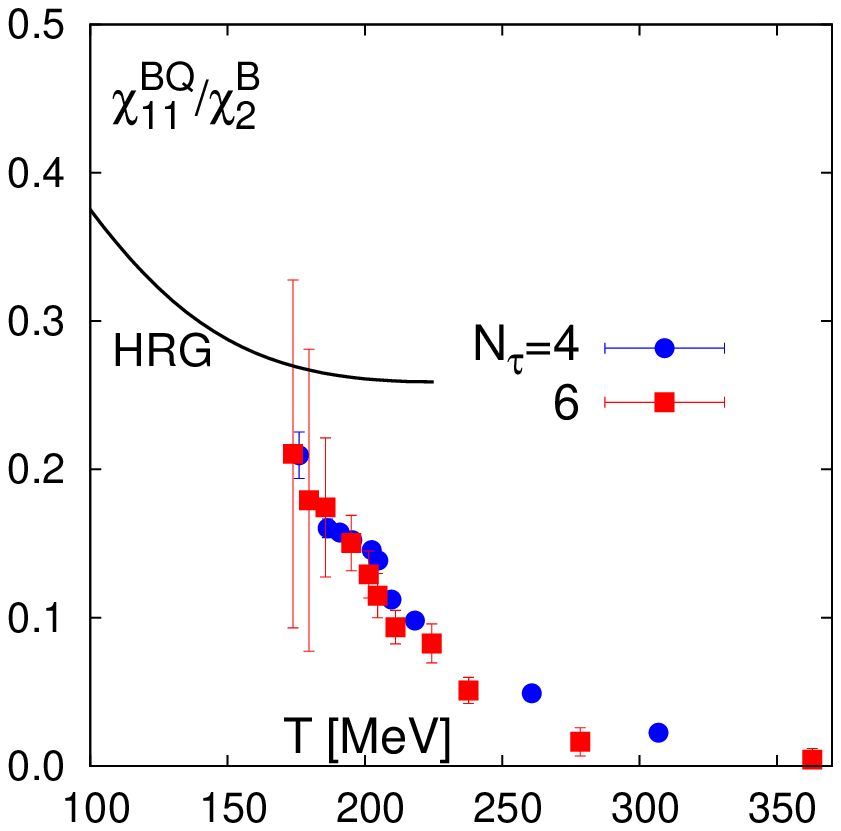}}
	\resizebox{0.32\textwidth}{!}{\includegraphics{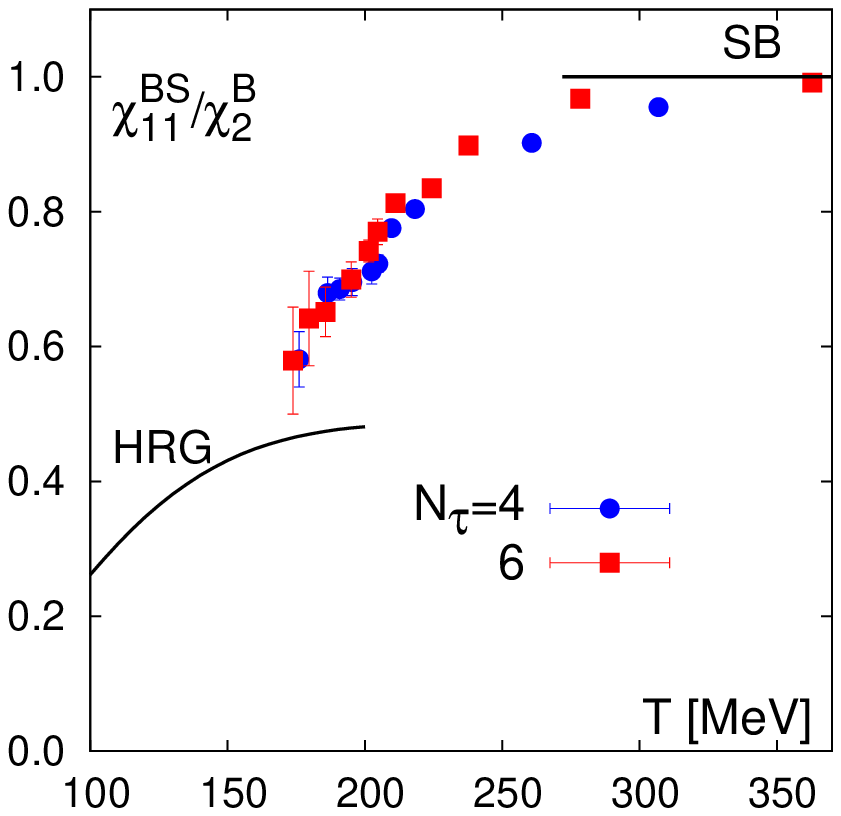}}
	\resizebox{0.32\textwidth}{!}{\includegraphics{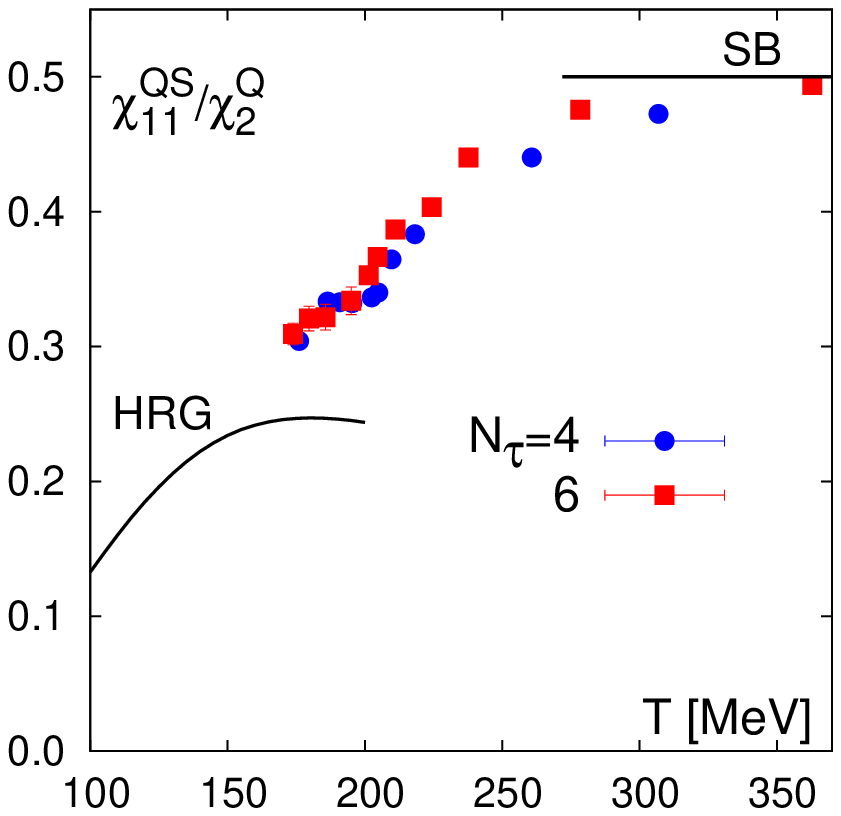}}
\end{center}
\caption{
Pairwise correlatios of the conserved charges baryon number (B),
electric charge (Q) and strangeness (S) as function of the
temperature, normalized to the quadratic fluctuations of B and Q
respectively.
}
\label{fig:R11BQS}
\end{figure}

In Fig.~\ref{fig:R11BQS}, we show the various correlations $\chi_{11}^{BQ}$,
$\chi_{11}^{BS}$ and 
$\chi_{11}^{QS}$ normalized to quadratic fluctuations $\chi_2^B$ and $\chi_2^Q$
respectively. The results from $N_\tau=4$ and $6$ lattices agree with each other
very well, and they are compared with the HRG model in the low temperature phase
and Stefan-Boltzmann limit in the high temperature phase.
We find that the correlations reproduce the qualitative behaviour of the HRG
model below $T_c$ and again start to agree with the free gas predictions
for $T\gsim 1.5T_c$. The rapid suppression of fluctuations above $T_c$ as well 
as the agreement of the correlations with the free gas predictions in the 
high temperature phase, suggests that in the quark-gluon-plasma the baryon
number and electric charge are predominantly carried by quasi-particles, with the 
quantum numbers of quarks \cite{EKR}.

\section{Conclusions}
\label{sec:con}
We have presented results on the equation of state on lattices of 
$N_\tau=4,6$ \cite{EoS,milc_eos} and $8$ \cite{hotQCDeos} obtained with two 
different kinds of improved staggered fermions. Hadron
masses have been kept constant in physical units and are chosen such
that we have a physical strange quark mass ($m_s$) and 2 light quarks
with a mass of $m_l=0.1 m_s$. We also presented some preliminary results
with smaller quark masses $m_l=0.05m_s$. We find that our two actions lead 
to a
consistent picture of the thermodynamics of QCD and find in particular
for the $N_\tau=8$ results only small cut-off effects. We have
calculated the equation of state as well as the velocity of sound and
find the softest point of the equation of state to be
$(p/\epsilon)_{min} \simeq 0.09$ at energy densities of $1-2$~GeV/fm$^3$.

Furthermore, we calculated corrections to the equation of state
arising from a non-zero baryon chemical potential, by means of a
Taylor expansion of the pressure. Using the expansion coefficients, 
 we have analyzed the quadratic and quartic fluctuations
of baryon number, electric charge and strangeness, as well  their
ratios. We find these quantities to be in good agreement
with the free gas results at temperatures of $T>1.5T_c$. Below $T_c$, 
qualitative features of the resonance gas are reproduced. The ratio
for the baryon number is closely related to the second approximation of 
the convergence radius of the Taylor series of the pressure with respect
to the baryon chemical potential. 

\section{Acknowledgments}
This work has been supported in part by contracts DE - AC02 - 98CH10886
and DE - FG02 - 92ER40699 with the U.S. Department of Energy.
Numerical simulations have been performed on the BlueGene/L computers
at Laurence Livermore National Laboratory (LLNL) and the New York Center
for Computational Sciences (NYCCS) as well as on the QCDOC computer of 
the RIKEN-BNL research center, the DOE
funded QCDOC at Brookhaven National Laboratory (BNL) and the apeNEXT 
at Bielefeld University.

\end{document}